# Nearly Century-scale Variation of the Sun's Radius

K. M. Hiremath[1], J. P. Rozelot[2], V. Sarp[3], A. Kilcik[3], Pavan D. G.[1], and Shashanka R. Gurumath[4]

[1] Indian Institute of Astrophysics, Bengaluru-560034, India; hiremath@iiap.res.in
[2] Université Côte d'Azur, Grasse-F-06130, France; jp.rozelot@orange.fr
[3] Akdeniz University Faculty of Science, Department of Space Science and Technologies, 07058, Antalya, Turkey
[4] Physical Research Laboratory, Ahmedabad, India

## Abstract

The Kodaikanal Archive Program (India) is now available to the scientific community in digital form as daily digitized solar white light pictures, from 1923 to 2011. We present here the solar radius data, obtained after a painstaking effort to remove all effects that contribute to the error in their measurements (limb darkening, distortion of the objective lens, refraction, other instrumental effects, etc.). These data were analyzed to reveal any significant periodic variations, after applying a multi-taper method with red noise approximation and the Morlet wavelet transform analysis. After removing obvious periodic variations (such as solar rotation and Earth annual rotation), we found a possible cycle variation at 11.4 yr, quasi biennial oscillations at 1.5 and 3.8 yr, and Rieger-type periodicity at ≈159, 91, and 63 days. Another ≈7.5 yr periodicity (as a mean) resulting from two other main periodicities detected at 6.3–7.8 yr can be identified as an atmospheric component. The detrending data show, over a mean radius of $959''.7 \pm 0''.7$, a residual of less than ≈(−)1 mas over the time period of analysis: if not spurious, this estimate indicates a faint decline, but probably confirms more the constancy of the solar diameter during the considered ranging time, within instrumental and methodological limits. The Kodaikanal long quality observations contribute to international efforts to bring past solar data measurements to the community to further explore issues, for instance, those of the luminosity/radius properties that could be used to pinpoint the "seat of the solar cycle."

Unified Astronomy Thesaurus concepts: Solar radius (1488)

## 1. Introduction

Due to the progress achieved in recent years in high angular resolution and in the precise determination of ephemerides, one might believe that the size of the Sun and its deviation to a perfect sphericity would be known at a very high precision, or at least as accessible as radii obtained nowadays on stars by interferometry. In spite of sophisticated techniques, the direct measurement of the roundness of the Sun and its possible temporal variability often remain vitiated by significant errors of measurements, both at the ground level (mainly due to atmospheric turbulence) and in space (limb shape displacements, space contamination of the devices, etc.). A statistical analysis allows us in principle to differentiate between different scenarios, but such a study has not been achieved so far in the absence of relevant data over a sufficient time span.

### 1.1. The Kodaikanal Archive Program (India)

The Kodaikanal Archive Program set up under the aegis of the Indian Institute of Astrophysics in Bangalore (India) is now available to the scientific community in digital form showing daily solar digitized white light pictures. These calibrated (without removal of limb darkening) images date from 1923 January 2 to 2011 August 2. They include many that can be judged to be of high quality by modern standards. The digitized images are hosted at http://kso.iiap.res.in. The Kodaikanal Solar Data Archive also provides other images, such as full- disk Ca K images, Ca K prominences, and $H_\alpha$ images. The solar white light images are analyzed here to estimate the Sun's radius. However, for other analysis, it is also possible to get access to the sunspots, their umbra and penumbra, as well as their area. With this nearly 89 yr scale data, the aim of the present study is to investigate long-term periodicity of the radius and, if possible, to determine whether or not it depends upon the solar activity.

### 1.2. Goal of the Study

The long standing question of the temporal solar radius variations is of interest for at least two reasons. The first one is that historical observations of the solar radius over the last 300 yr may provide an estimate of the past luminosity changes and their possible climate effects on the Earth and its environment (see, for example, Hiremath & Mandi 2004, Hiremath 2009, Hiremath et al. 2015, and references therein). Second, a possible shrinking or expanding global shape of the Sun is directly linked to the gravitational effects in the nearly subsurface layer (NSSL, called leptocline; see Rozelot et al. 2016; Kosovichev & Zhao 2016 Kosovichev & Rozelot 2018a), which is of importance to understand the role of the underlying magnetism.

Indeed, although the ultimate source of the solar energy is provided by nuclear reactions taking place in the center of the Sun, for which the rate is certainly constant on timescales shorter than millions of years, the immediate source of energy is the solar surface. Observations of the solar radiation integrated over the entire solar spectrum (total irradiance) are known to vary on timescales of minutes to the 11 yr solar cycle. Thus, if the central energy source remains constant while the rate of energy emission from the surface varies, there must be an intermediate reservoir where the energy can be stored or released depending on the variable rate of the energy transport. The gravitational field of the Sun is one such energy reservoir. If the energy is stored in this energy reservoir, it will result in a change in the solar radius (Pap et al. 1998, 2001; Fazel et al. 2008). Therefore, a careful determination of the time dependence of the solar radius can provide a constraint on models of total irradiance variations, either on a short-term time range or a longer one. To this respect, there is still a lack of coherent measurements on a century scale.

Measurements of the solar radius have been carried out since the highest Antiquity, but it was not until the 17th century that they became reliable under the impetus of the French astronomer Picard. Since then, even with gradually increasing accuracy by different means or techniques, the results obtained so far do not allow us to determine without ambiguity its estimation in absolute value, its variation, or its constancy as a function of solar activity. Reviews of the existing data have been made, for example, by Wittmann (1977) from 1836 to 1975 or by Djafer et al. (2008) from 1981 to 2006; a more complete history, can be found in Rozelot & Damiani (2012), at least up to the year 2011, and upgraded in Rozelot et al. (2016).

If there is no consensus today on the abovementioned issues, this could be due to several reasons: (i) there is no unique method of estimation of the Sun's radius; (ii) instrumental effects (point-spread function, spatial resolution, optical distortion, and temperature variations) are not properly taken into account; (iii) atmospheric turbulence (from the ground) on the observed Sun's full-disk pictures or on the Sun's limb, which are not tidily corrected; (iv) different instruments are observing at different phases of the solar cycle; and (v) observations are not made at the same wavelengths. In regards to this last issue, certain instruments are observed in the continuum at different wavelengths, while others are observed in the center of a Fraunhofer line, as was done at the Mount Wilson observatory (USA). In addition, there are instruments that used a narrow bandpass, such as MDI on board Solar and Heliospheric Observatory (SOHO) or Mount Wilson magneto- grams whereas others use a wide spectral domain on the order of hundreds of nanometers, as was done by the CCD astrolabes.

In regards to the phase of the radius with solar activity, broadly speaking, it can be noted that at least four groups of observers have claimed that the solar radius varies in phase with surface activity, seven groups of observers have reported radius changes in antiphase with surface activity, while four groups of observers have reported no significant change at all (see Stothers 2006 and references therein). From space, Kuhn et al. (2004) have reported that during solar cycle 23, between solar minimum and solar maximum, the radius of the Sun did not change by more than ± 7 mas (14 mas peak to peak). From f-mode helioseismology, the so-called seismic radius varies in antiphase with solar activity with a different amplitude according to the cycle (Kosovichev & Rozelot 2018a). Two radius estimates must be distinguished: the physical (or true) radius and the seismic (or acoustic) radius. The former has been widely used in records dating back to the eighteenth century (Vaquero et al. 2016, and references therein), and is still widely used, while the concept of seismic radius has only been in force for a couple of decades (Schou et al. 1997), the link between the two can be made only through modeling. The seismic radius was adopted as a standard by the International Astronomical Union in 2015.

The paper is organized as follows. In Section 2, we present the data used and the method of analysis. Results are presented in Section 3, including the periodic analysis, and Section 4 ends with conclusions.

## 2. Data and Methodology

Established in 1899 as the Solar Physics Observatory, the Kodaikanal Observatory has since been a leading observatory where solar observations at different wavelengths have been continuously archived. The observatory is in South India (Tamil Natu State) at an elevation of 2343 m. The meteorological conditions are in general quite good, allowing on average around 280 days of observations per year. Recently, ≈89 years of data of white light images were digitized to be available to the scientific community, as mentioned in Section 1.1. The solar telescope at the Kodaikanal Observatory consists of a 15 cm achromatic lens as the objective, which has a focal length of 240 cm. The telescope produces a solar disk image of about 20 cm in diameter. For the period of time between 1923 and 2011, a total of 24,939 images are available to be used for the present investigation, out of which around 23,200 are of very good quality. Details of the telescope (Sivaraman et al. 1993) and the digitization process of the images can be obtained, for example, in Ravindra et al. (2013) or Priyal et al. (2019).

One of the major steps in our investigation is the computation of the center and radius of the solar disk from the images. Many methods can be used. The Circle Hough Transform here chosen by Ravindra et al. (2013) using a Sobel filter isolates the sharp edges in the solar images. This procedure not only detects the solar limb, but also detects the borders of the features and sharp gradients. The calibrated digitization process using such a code pays particular attention to the removal of the limb darkening for all the images, as well as active regions (spots and faculae) at the limb. After correctly detecting the edge of the solar image, by considering all the detected pixels of the limb, for all the solar daily images, a circle is fitted so that the center of the solar image and its radius, with its uncertainty, are simultaneously and uniquely computed. Depending on the quality of the images, the error on the individual computed radius may vary from 0 02 (as a minimum) to 0 3(0) (as a maximum; larger ones are discarded, see further). Mathematical details of such calculations are presented in Appendix A (see also previous studies by Pucha et al. 2016 and Hiremath et al. 2019).

Large sunspots near the limbs reduce the estimated radius of the image. Each daily image was manually examined and images with sunspots near the limb were rejected. A consistent criterion was also adopted in such a way that if the deviation from the average radius is greater than nearly one standard deviation, such images are not considered for the radius estimation. This resulted in reduction of the data (for example, in the case of the year 1960, the reduced data is only about six months; in order to maintain the continuity of the yearly averaged radius, this half year data set was also considered). However, we find that this reduction of such a data set has not introduced any bias in estimation of the solar radius and is within the error bar of grand average solar radius. Uncertainty of estimated radius in the daily image is computed as follows. For the daily solar images, we find, on average, an estimated uncertainty of ≈0".2(3) (except during the occurrence of sunspots near the limb, for which the uncertainty would be higher, but they have been removed). For the time being, our analysis does not permit us to check effects with the heliographic latitudes (departures to sphericity).

Lastly, it is necessary to calculate the daily values of heliographic latitude ($B_o$) and longitude ($L_o$) of the disk center as well as the polar angle P which has been made through the heliographic latitude θ, heliographic longitude L, and longitude difference from central meridian l. Correction for distortion of the Sun's image has also been considered as a telescope objective lens with a short focal length as in Kodaikanal's observatory may contribute to a certain distortion of the projected image. This distortion was corrected by using an empirical relation (Beck et al. 1995), as explained in Appendix B. Although it is very complicated to estimate and disentangle the effect due to distortion, first order distortion may contribute to ≈0".3 in the final result (note that dominant contribution due to radial distortion can be expanded in terms of Taylor series but higher terms will be negligible). It must also be noted that during the investigated period of time (1923–2011), at the Kodaikanal Observatory, imaging technology was not changed, the same instrumentation was used, and the observed instrument remains in the same place.

The left panel of Figure 1 illustrates the number of images used for the present investigation by considering daily data; the right panel shows the monthly Fried coherence parameter ($r_0$), in cm, for which the mean value is 4.8 ± 1.1 cm. It is not uncommon to get a seeing of 6–8 cm. On most days only one image is obtained and on some days a few more images are obtained, depending on the weather conditions. Most of the observations are taken early in the morning, before 10 am (IST), for which the Fried parameter hovers around 4–6 cm; around 2% of the observations can be associated with a Fried parameter of up to 10 cm (see note[5]). According to observational seeing standards, the Kodaikanal site can be qualified as being of good quality.

---

[5] Unpublished results by Dr. Sridharan from the Indian Institute of Astronomy (IIA) in Bangalore (I) –personal communication.

## 2.1. Radius Estimation from the Data

The collected archived data need to be corrected and cleaned. We first apply the refraction corrections. For zenith distance up to 70°, the refraction integral can be evaluated with good accuracy without any hypothesis about the structure of the atmosphere, as it depends only on the air index, temperature, and pressure at the observational site during the data recording (for a complete analysis, see, for instance, Chollet 1981). This issue justifies that a large number of nearly equivalent approximate formulas have been derived that do not require the full knowledge of the structure of the real atmosphere: see, for instance, Wittmann (1977) or Meeus (1999) for more details. Numerous theories of astronomical refraction (i.e., for objects outside the Earth's atmosphere) have been proposed, and a great number of algorithms for its practical computation have been put forward. Elaborated algorithms or detailed refraction tables are available and can be useful. Lastly, Ikhlef et al. (2019) made a comprehensive review of this issue for modern solar astrometric measurements. As all these algorithms give nearly the same results, we used here the Meeus formulas which are quite satisfactory. The asymmetric refraction, also called abnormal refraction, regards the part that cannot be explained by the quoted former models (Teleki 1979). Contrary to symmetric-radial refraction, abnormal refraction is dependent on both zenital distance and azimuth. Chollet (1981), in a study that stands as a corner stone for later works, denotes that it is not necessary to count for it, at least at the level of accuracy given by the "normal" refraction. Then the corrected apparent measurement of the solar radius is carried back to 1 au according to the Ephemerides. We first used the "NASA JPL Horizons ephemeris system." This service does not list Kodaikanal in their predefined file of observatories, but it allows us to use coordinates of the observatory manually. The ephemerides from the "Institut de Mécanique Céleste et de Calcul des Éphémérides" allow also to compute the Sun–Earth distance for the Kodaikanal coordinates, that can be compared with the NASA ephemeris. The differences are about E-05 of astronomical units. The corrected radius data shows a long (positive) shift over the ranging time under consideration, certainly an instrumental effect on the long term, that we were unable to identify. However, the last part of the data, after the year 2000, do not present such a shift and moreover are more accurate, certainly due to the instrumentation quality, which was at the best. We choose this part of the data as a reference to detrend the other data, getting a mean radius value of 959″.7 ± 0″.5.

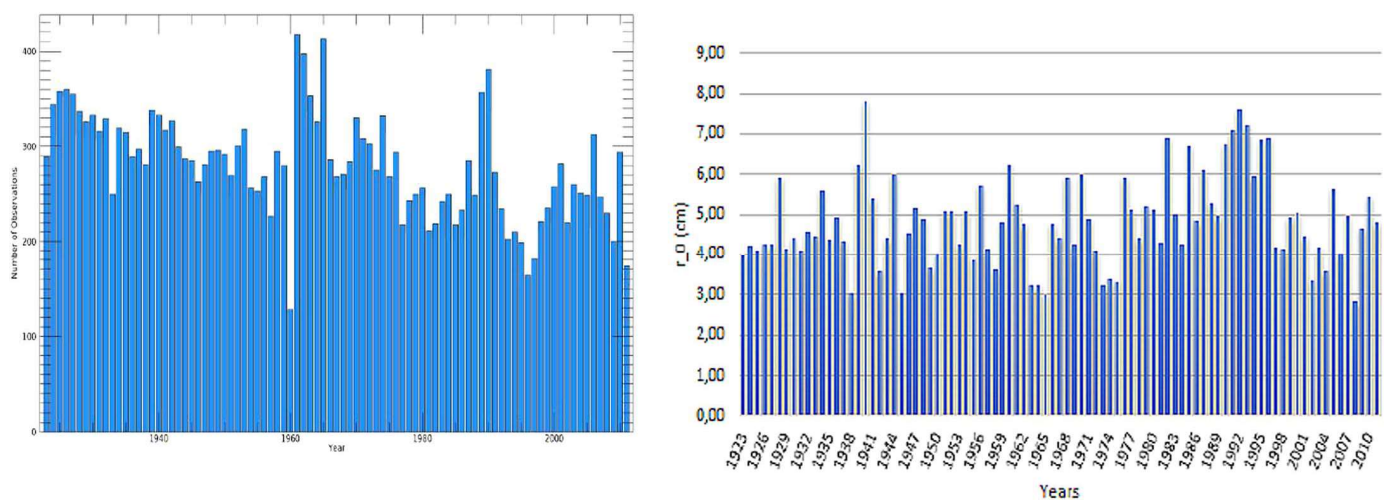

Figure 1. White light observations of the solar disk, made at the Kodaikanal Observatory (South India), from 1923 January 2 to 2011 August 2. Left: number of observed white light pictures sorted by month. A total of 24,939 images have been kept for this study. There were no observations from 1960 June 17 to December 31. Right: annual Fried coherence parameter ($r_0$, in cm) during the time of solar observations. The mean value is 4.8 ± 1.1 cm. The Kodaikanal seeing can rise up to 8–10 cm on certain days; the site can be qualified as being of good quality.

Results are shown in Figure 2. The two horizontal lines up and down show 1σ and 2σ levels. For the sake of clarity, the error bars have not been drawn; the mean standard deviation (over the whole data) is 0″5. We found that 1σ discards 28.3% of the measurements (7063) and 2σ discards 4.3% (1064). For the following analysis, we used the 2σ level to remove outliers.

## 2. Analysis of the Data and Results

Figure 2 exhibits pseudo-periodic oscillations. In order to quantify them, we first applied the MTM (Thomson 1982). This method is a quite powerful tool for determining low amplitude harmonic oscillations with a high degree of statistical significance over the full duration of each of the investigated time series. This issue is due to the fact that to provide spectral estimates the method uses both the line components and the continuous background of the spectrum. The technique uses orthogonal windows (or tapers) to obtain approximately independent estimates of the power spectrum and then combines them to estimate the power spectral density (Ghil et al. 2002). MTM analysis was extensively used to analyze various solar parameters and to detect their periodic properties (see, for instance, Kilcik et al. 2016, 2018, and references therein). Here, three sinusoidal tapers were used meaning that the number of the peak is changing from one to three (see Ghil et al. 2002, Figure 13), the data being analyzed for all frequency intervals. The detected signals were accepted when a 95% red noise confidence level was reached.

The MTM enables us to obtain the exact values of the periods, but it does not give any information about the time over which the periodic behavior extends (or in other words the temporal variations of the found periods). To remove this uncertainty and check the existence of such periods, we applied a Morlet wavelet analysis with a red noise approximation. The Morlet wavelet method is a powerful tool for analyzing localized power variations within a time series (Torrence & Compo 1998). This method has also been extensively applied in solar physics (see also Kilcik et al. 2016, 2018). By contrast to the classical Fourier or MTM analysis, the Morlet wavelet transform uses wavelets characterized by scale (frequency) and time localization. The main gain of the method is that it enables us to show varying periods with time. The Morlet wavelet is a complex sine wave, localized within a Gaussian window (Morlet et al. 1982). Its frequency-domain representation is a single symmetric Gaussian peak, and its localization is very accurate. The use of such a wavelet has the advantage of incorporating a wave with a clear period that is finite in extent. The Kodaikanal radius data were analyzed using the standard Interactive Data Language package for Morlet wavelet analysis, and the scalograms were obtained to study both the presence and evolution of the periodicity. This analysis was carried out using the red noise approximation with 90% confidence, using the monthly data over the investigated time period.

### 3.1. Results

Solar magnetic activity characterized by proxies such as sunspot number, sunspot areas, solar flares, radio bursts, neutrino flux, etc., exhibits a number of periodicity. In addition to the well-known ≈11 yr Schwabe cycle (which may range from 98 to 156 months), the most known are the quasi biennial oscillations (QBOs) and the Rieger-type periodicity (RTP). The first one appears between 1.5 and 4 yr (18–48 months). Solar dynamo is thought to be the mechanism responsible for the generation of these QBOs (Inceoglu et al. 2019). The second type includes the Rieger periodicity, first observed at 155 days (Rieger et al. 1984), ranging now in literature from 150 to 160 days, and other periodicity detected at 128, 102, 78, and 51 days. Although there is presently no solid ground for a theoretical explanation of the RTP, the most appropriate mechanism suggested would be of Rossby-type waves (Dimitropoulou et al. 2008). It was also argued that periodicity could originate in strongly magnetized regions, related to the timescale taken for the storage and/or escape of the magnetic field in the solar convection zone (Ichimoto et al. 1985). Hiremath (2010) proposed that the observed quasi-periodicity of solar activity indices in the range of 1–5 yr could be due to Alfvén wave perturbation of the strong toroidal field structure and variation of very long period solar cycles. Solar radius periodicity has not really been searched for up to now, essentially due to a lack of long reliable series that can be used unambiguously. The Kodaikanal series provided the opportunity to use novel and powerful analyses to detect possible oscillations.

Results from the MTM Analysis in Figure 3(a) shows the power spectra obtained with the help of the MTM period analysis. It exhibits periodicity at 128–146, 98–108, 76–82, 43–47, 17.8, 11.2–12.5, 5.9–6.1, 5.3, 4, 3.0–3.2, 2.6, and 2.0–2.2 months. Small gaps in the data have been filled with cubic spline interpolation and one 6 months gap is filled with linear interpolation. Indeed, due to the lack of observations during the year 1960, between June 17 and December 31, we first made an analysis of the data in two parts: the first one from the beginning of the observations up to the beginning of the gap, and the other one from the end of the gap up to the end of the data. Results do not show significant differences with the analysis of the full data, as presented here. The three curves show the 90%, 95%, and 99% confidence levels respectively.

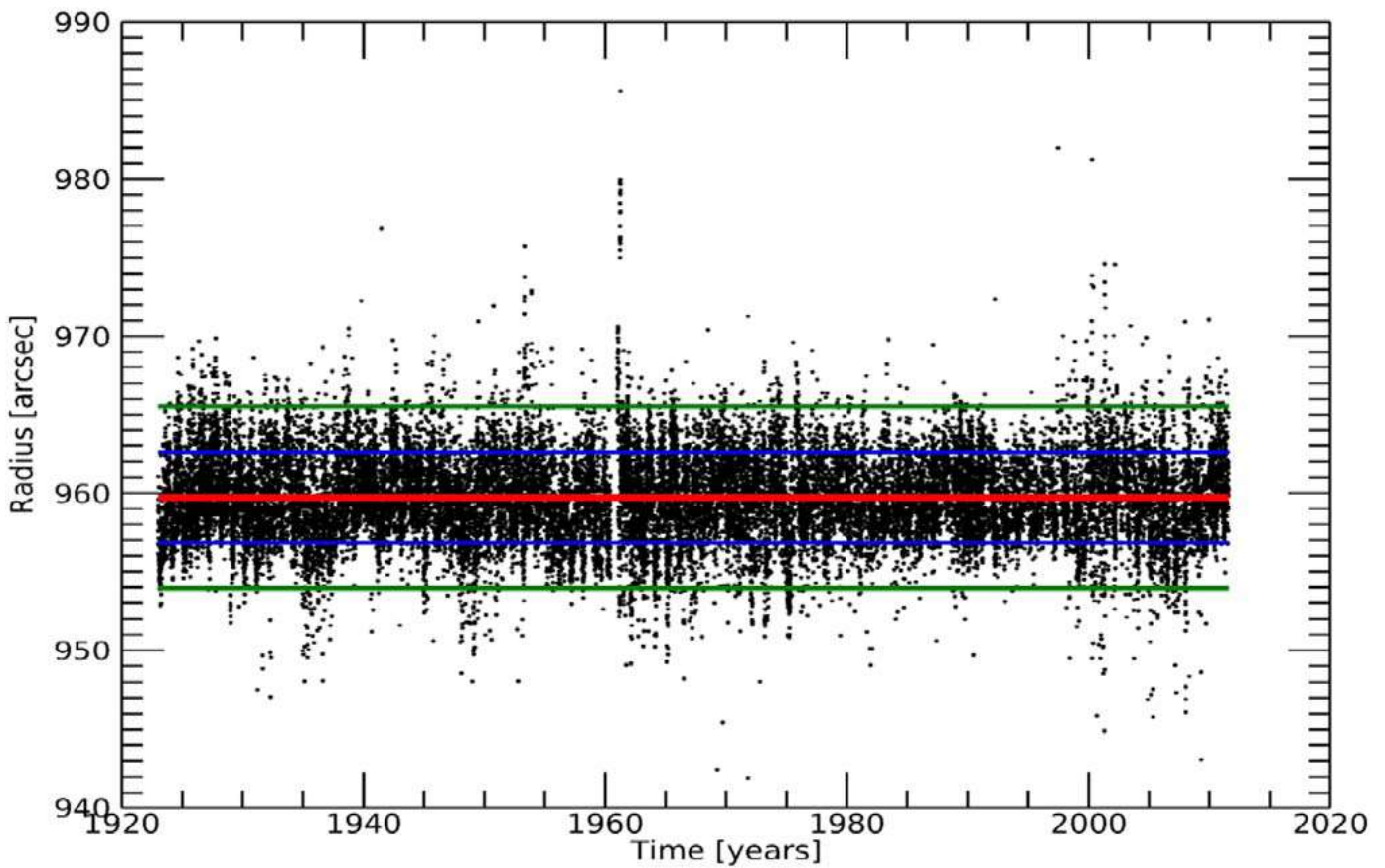

Figure 2. Daily solar radius data as obtained at the Kodaikanal observatory (South India), carried back to 1 au, after refraction corrections and after removing a long-term (positive) instrumental shift. The mean radius is 959". 7. The two horizontal lines (up and down) show 1σ and 2σ levels. For the sake of clarity, the error bars have not been drawn; the mean standard deviation (over the whole data) is 0 5. The intermediate line shows the very faint negative linear trend.

Results from the Morton analysis in Figure 3(b) show the results obtained when applying the Morlet wavelet analysis. The effect of edges is represented by the cone of influence (COI) plotted with hatched lines. In this case, the periods detected above the 90% confidence level inside the COI are shown by thin black contours. This plot enables us to see how the detected periods evolve with time. Interpretation of the observed periodicity is quite obvious and can be listed as follows:

1. Periodicity at 12 months (11.2–12.5) found in the MTM plot reflects the annual variation; it clearly appears in the wavelet plot around the years 1930 and 1960; the periodicity at 4 months (120 days) is clearly an harmonic.
2. Periodicity at 2.6 months found in the MTM plot is a possible harmonic of the solar rotation occurring at 27.2 days (ratio of ∼3.0); this period is also appearing in the wavelet scalogram without any time preference.
3. Periodicity at 17.8 months is a QBO; it appears strongly around the year 1980 and weakly around the years 1995 and 2005 in the wavelet scalogram.
4. Periodicity at 43–47 months (within a 1.1 month error) can be identify also as a QBO; it appears strongly around the years 1950–1960 and weakly around 1995 in the wavelet scalogram.
5. Periodicity at 5.3 months (159 days) found in the MTM plot is due either to the seasonal variations or likely is the appearance of the 155–160 days Rieger periodicity. This periodicity does not have any time preference and appears everywhere in the wavelet scalogram.

The two following periodicities detected at 2.1–2.2 months (63 days) and at 3.0–3.2 months (92 days) can be identified as likely being the same as the well-known 51 and 102 days RTP (considering the uncertainties), unless they are harmonics of the solar rotation. The 2.1–2.2 months is appearing in the wavelet scalogram without any time preference.
Lastly,
1. Periodicity at 76–82 months (within a 3.3 months error; 6.7 yr as a mean) and 98–108 months (within a 5.7 months error; 8.6 yr as a mean) can be due to atmospheric fluctuation components as reported by many authors since the pioneering work of Cimino ([1944](#)); they appear around the years 1930–1940 and 1945–1965 in the wavelet scalogram.

2. An about 11.4 yr solar cycle periodicity is detected (128–146 months within about 10 months error); it appears strongly around 1930–1945 but this periodicity is outside the COI in the wavelet scalogram.

The Calern team (Caussols Observatory, South France), by means of a solar astrolabe instrument, found a mean oscillation at ≈900 days of ≈0 5 of amplitude (Delache et al. [1985](#)), in phase opposition with the solar cycle. This issue was the source of many commentaries and papers. In spite of efforts to reconcile different measurements, made all around the world, by different means but all aiming to consolidate this radius oscillation, it was not able to be confirmed or denied: "the origin of these apparent variations is currently unclear" (Morand et al. [2011](#)). We do not find this periodicity in our results that may confirm our previous findings (Rozelot & Damiani [2012](#)): the 7.5 yr atmospheric component is a quasi-harmonic (3) of that found in the Calern data as 7.5 × 365 (days) = 2737, divided by ≈900=3.04). Delache et al. ([1994](#)), analyzed astrolabe measurements of the solar radius obtained at Calern Observatory by means of a Fourier transform. The daily values have been smoothed by a running linear filter over a three-month window. They obtained characteristic peaks at 136.6, 61.6, 30.5, 20.4, 15.6, 13.4, 11.5, 10.4, 9.5, and 8.3 months. We can notice some similarities with our results. The two last periodicities are the same as those already found in the radius spectra of the solar observations made at the Campidoglio Observatory (Roma, I) and identified as atmospheric frequencies (Giannuzi [1953](#), [1955](#)).

### 3.2. Near Century-scale Variations of the Solar Radius

After removing the instrumental (positive) shift, an apparent feeble decreasing trend on a century scale appears, of

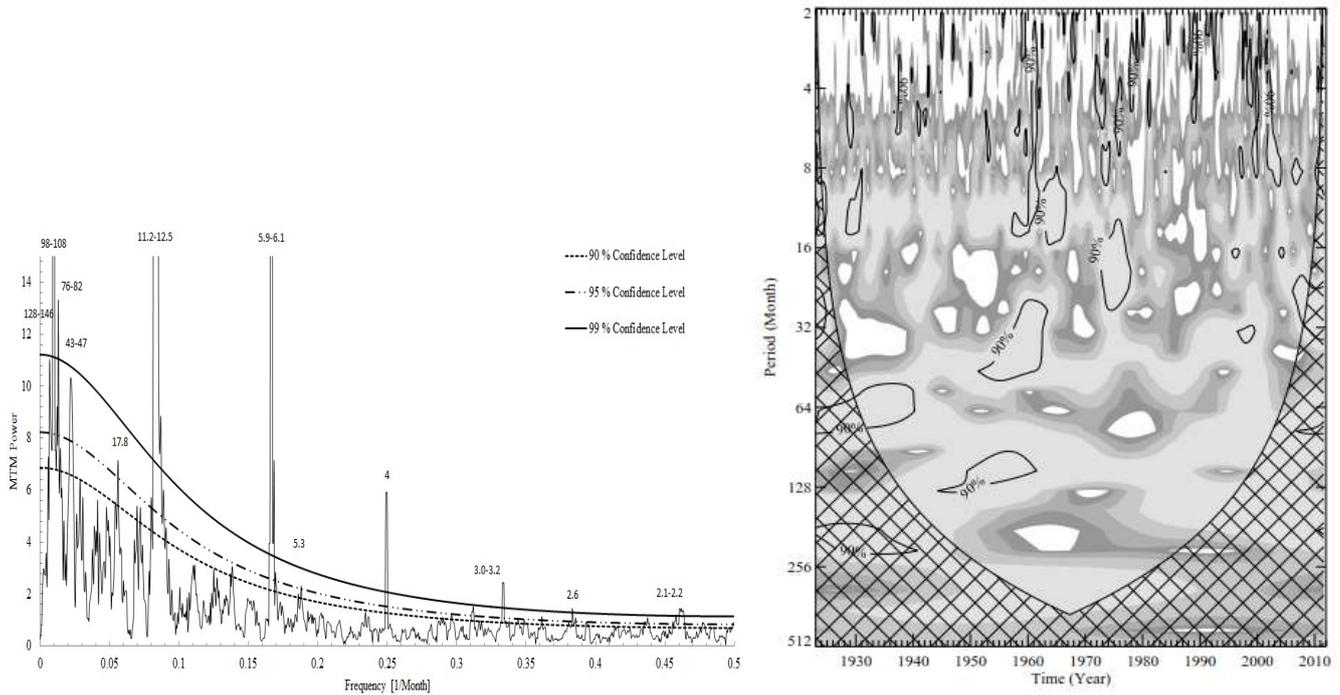

Figure 3. Periodic analysis performed over the daily Kodaikanal radius data, from 1923 to 2011. (a) Multi-taper method (MTM). The lines are drawn at 90%, 95%, and 99% confidence levels. Numbers at the top of the peaks are given in years. (b) Morlet wavelet method. The black contours in the wavelet scalogram indicate a 90% confidence level and the hatched area below the thin black line is the cone of influence (COI).

≈−1.1 mas per year (± 0.4 mas). If not spurious, this trend is far less than those found for the first time by Eddy & Boornazian ([1979](#)) but of the order of magnitude found by the other authors listed in Table 1 in Kuhn et al. ([2004](#)) or even those obtained more recently by Bush et al. ([2010](#)).

Observations of long-term changes in solar diameter have already been reported by different means. From the Greenwich (UK) measurements of the solar diameter from 1836 through 1953, a secular change of −0.023 arcsec yr$^{-1}$ has been found by Eddy & Boornazian ([1979](#)), but such a large rate has been questioned. Shapiro ([1980](#)) analyzed observations of 23 transits of Mercury in front of the Sun between 1736 and 1973 and found a decrease of 0.003 arcsec yr$^{-1}$. Reconsidering meridian circle observations with analysis of transits of Mercury and durations of total solar eclipses, Parkinson et al. (1980) indicated a secular change of −0.0014 arcsec yr$^{-1}$ (± 0.0008) over 250 yr of observations. Measurements of the solar diameter made during the 1981 July 31 total eclipse, together with a reanalysis from 1715 to 1981 of previous eclipse data (110 timings) and Mercury transits (2150 timings) led Parkinson (1983) to induce that if there is no strong evidence for any secular change in the solar diameter there is increased support for an approximately 80 yr cyclic variation. Such a conclusion has been widely discussed by Gilliland (1981), who concluded that a secular decrease of less than −0.0015 arcsec yr$^{-1}$ is likely from an objective analysis of the data. Such a value is fully consistent with the value of −0.0013 ± 0.0008 arcsec yr$^{-1}$ obtained by Dunham et al. (1980) who analyzed three solar eclipses in England in 1715, in Australia in 1976 and in North America in 1979. Lastly, Morrison (quoted by Shapiro 1980), after a reanalysis of observations of transits of Mercury reported a secular change of −0.0014 (±0.0008) arcsec yr$^{-1}$ over the 250 yr interval from 1723 to 1973. Hence converging fasces of results lend credence to the idea that, irrespective of solar activity, the Sun's radius almost remains constant (at least at a level of under 0 01 per century) over the nearly century scale studied.

### 3.3. Does Solar Radius Correlate with Solar Activity?

A scatter plot drawn between the sunspot number and the radius data does not show any clear correlation. Considering that our full analysis is somewhat limited by the uncertainties of the atmospheric seeing, we can conclude that over a century scale, the solar radius remains nearly constant and is quasi-independent of changes in the magnitude of the solar activity. Such results are consistent with previous studies, conducted from the ground, such as those of Labonte & Howard (1981) and Lefebvre et al. (2006) deduced from data obtained at Mount Wilson Observatory (USA), those of Neckel (1995) obtained at the National Solar Observatory at Kitt Peak (USA), and those of Wittman & Bianda (2000) at Locarno (CH) and Izaña (SP). Analyzing 250 yr (1773–2006) of solar radius data monitored at the Royal Observatory (ROA) at Cadix (SP), Vaquero et al. (2016), came to the same conclusion that the Sun's radius remains quasi constant, within instrumental and methodological limits, and showing no relationship with the sunspot number index, at least at one standard deviation. Such results, all obtained from ground-based instruments must be nuanced by those obtained from space. Indeed, space observations are in principle of higher accuracy, as they are free from atmospheric disturbances. For example, with the help of the MDI–SoHO instrument, Emilio et al. (2000) were able to put an upper limit on possible secular variations of the radius fluctuation measurements, of 8.1 ± 0.9 mas yr$^{-1}$ together with a solar cycle radius variation of less than 21 ± 3 mas (peak to valley). Revisiting the results, the same authors (Bush et al. 2010) claimed that "any intrinsic changes in the solar radius that are synchronous with the sunspot cycle must be smaller than 23 mas peak to peak, and it must not be changing (on average) by more than 1.2 mas yr$^{-1}$". Meftah et al. (2015) found, for the years 2010–2011 and through data obtained by means of the PICARD satellite, that changes in solar radius amplitudes that were less than ≈20 mas, indicating a very faint link with the solar activity. Lastly, space data from MDI (on board SoHO) and HMI (on board of SDO), show with a very high accuracy that the solar seismic radius is varying with the solar activity, in a nonhomologous way, in the NSSL. The seismic radius was found to be reduced by about 3 km during the Cycle 23 maximum in 2000–2003, and by about 2 km during the Cycle 24 maximum in 2013–2015; the strongest changes occurring at the depth of ≈5.2 Mm (Kosovichev & Rozelot 2018b). In our view, as ground-based instruments are not affected by degradation due to space environment, and as maintenance can be easily provided, the faint variations, as reported above from space instruments, could be detected only if the atmospheric effects were properly monitored and taken into account.

### 4. Conclusions

The main goal of this work was to measure the solar disk radius over a long period of time. However, the techniques used in the past were not as sophisticated as they are presently; we do not have a corpus of calibrated systems for the instruments used at that time and no consistent statistics are available. Thus, it is difficult to remove systematic errors that we do not know, but that we could better considered today with modern methods. The digitized images present such systematic errors, mainly due to the distortion induced by the several processes at work, from the beginning (the recorded solar image) up to the final product, the radius. Long-term systematic errors are important in ground-based experiments and both long- and short-term systematics are important. Statistical errors decrease with the square root of the number of observations, but the systematics do not go away. The best careful analysis of the Kodaikanal's observations has been done here, but we are aware, concerning ground-based observations that the selected 2σ level might not be enough, albeit coherence of the results obtained with the MTM method gives credence in the analysis.

Many authors reported and discussed variations observed in the Sun's diameter measurements performed during the last three centuries (see, for example, Toulmonde 1997; Rozelot & Damiani 2012, Vaquero et al. 2016, Rozelot et al. 2018). Our results, deduced from the analysis of digitized white light images taken over a nearly century timescale of observations at the Kodaikanal Observatory, added greater interest to the solar diameter measurements and their long-term variations. We found typical periodicity in the data: quasi biennal oscillations at 1.5 and 3.8 yr and Rieger-type oscillations at 159, 92, and 63 days. The solar cycle was detected at 11.4 yr but a sunspot correlation of only −0.05 is visible. It was also found that the Sun's radius remains almost constant, showing a very feeble decrease over the century studied. This consistency of the solar radius in time remains an important topic: obtaining an appropriate long quality database is of interest to further explore issues on the luminosity/radius properties which could be used to pinpoint the "seat of the solar cycle."

**Acknowledgements.** The authors are grateful to the observers who have completely dedicated their lives to taking and archiving pictures of the Sun. They also thank all the people who were involved in the digitization project. K.M.H. is thankful to Prof. Jagdev Singh and Prof. Ravindra for useful discussions. J.P.R. would like to express his gratitude to the Indian Institute of Astrophysics in Bangalore for the several stays he had there as a visiting scientist. The authors also thank the referee for very helpful comments.

## Appendix A Circle Fitting

All the detected pixels on the edge of the image are least- square fitted with a circle that uses a system of three simultaneous equations with three unknowns to get a unique solution for center and radius of the solar image. This method only requires the edge of the circle coordinates to be the input and an initial guess is not required. This method uniquely computes the three required coordinates (two coordinates for center of the circle and the radius). The process is described in detail below:

If $x_i$ and $y_i$ (where $i = 1, N$, $N$ is total number) are Cartesian coordinates of the detected pixels and $\bar{x}$ and $\bar{y}$ are their respective means which are defined as follows

$$\bar{x} = \frac{\sum_i x_i}{N}, \quad (1)$$

and

$$\bar{y} = \frac{\sum_i y_i}{N}. \quad (2)$$

Let $x_i$ and $y_i$ be further transferred into new variables $u_i, v_i$ such that

$$u_i = x_i - \bar{x}, \quad (3)$$

and

$$v_i = y_i - \bar{y}. \quad (4)$$

Let $(u_c, v_c)$ be the center coordinates of the circle with radius $R$ and $\alpha = R^2$.

Distance of any point $(u_i, v_i)$ from the center is $= [(u_i - u_c)^2 + (v_i - v_c)^2]^{1/2}$.

From the method of least-square fit which implies function $S = \sum_i [g(u_i, v_i)]^2$ should be minimum where

$$g(u_i, v_i) = (u_i - u_c)^2 + (v_i - v_c)^2 - \alpha. \quad (5)$$

Hence, partial derivatives of these functions with respect to $\alpha$, $u_c$, and $v_c$ should all be zero. For the partial derivative of $S$ with respect to $\alpha$ we get

$$\frac{\partial S}{\partial \alpha} = 2 \sum_i g(u_i, v_i) \frac{\partial g}{\partial \alpha} = 0, \quad (6)$$

$$\Rightarrow -2 \sum_i g(u_i, v_i) = 0, \quad (7)$$

$$\Rightarrow \sum_i [(u_i - u_c)^2 + (v_i - v_c)^2 - \alpha] = 0, \quad (8)$$

$$\Rightarrow \sum_i u_i^2 + \sum_i v_i^2 + \sum_i u_c^2 + \sum_i v_c^2$$
$$- 2 \left[ \sum_i u_i u_c + \sum_i v_i v_c \right] = \sum_i \alpha, \quad (9)$$

$$\Rightarrow \sum_i u_i^2 + \sum_i v_i^2 + N[u_c^2 + v_c^2]$$
$$- 2 \left[ u_c \sum_i u_i + v_c \sum_i v_i \right] = N\alpha. \quad (10)$$

With the known fact that $\sum_i u_i = \sum_i (x_i - \bar{x}) = N\bar{x} - N\bar{x} = 0$ and $\sum_i v_i = 0$, we get

$$\Rightarrow \sum_i u_i^2 + \sum_i v_i^2 + N[u_c^2 + v_c^2] = N\alpha. \quad (11)$$

For the partial derivative of $S$ with respect to $u_c$ we get

$$\frac{\partial S}{\partial u_c} = 2 \sum_i g(u_i, v_i) \frac{\partial g}{\partial u_c} = 0, \quad (12)$$

$$\Rightarrow \sum_i (u_i - u_c) g(u_i, v_i) = 0. \quad (13)$$

On expansion, we get

$$\Rightarrow \sum_i u_i^3 + \sum_i u_i v_i^2$$
$$- 2u_c \sum_i u_i^2 - 2v_c \sum_i u_i v_i - u_c \sum_i u_i^2 \quad (14)$$

$$- u_c \sum_i v_i^2 - Nu_c^3 - Nu_c v_c^2 + N\alpha u_c = 0 \quad (15)$$

By substituting the value $N\alpha$ from Equation (11), we get

$$u_c \sum_i u_i^2 + v_c \sum_i u_i v_i = \frac{1}{2} \left[ \sum_i u_i^3 + \sum_i u_i v_i^2 \right]. \quad (16)$$

Lastly the partial derivative of $S$ with respect to $v_c$ yields the following equation

$$\frac{\partial S}{\partial v_c} = 2\sum_i g[u_i, v_i]\frac{\partial g}{\partial v_c} = 0. \qquad (17)$$

Following the derivations of Equations (11)–(16) we obtain the following equation:

$$u_c \sum_i u_i v_i + v_c \sum_i v_i^2 = \frac{1}{2}\left[\sum_i v_i^3 + \sum_i v_i u_i^2\right]. \qquad (18)$$

Simultaneous solution of Equations (16) and (18) yield the two central coordinates $u_c$ and $v_c$. Then from Equation (11), we get the value of $\alpha$ ($=R^2$) and, hence the radius $R$:

$$R^2 = (u_c^2 + v_c^2) + \frac{1}{N}\left[\sum_i u_i^2 + \sum_i v_i^2\right]. \qquad (19)$$

By adding two central coordinates $(u_c, v_c)$ to the respective means, original central coordinates $x_c$ and $y_c$ are obtained

$$x_c = u_c + \bar{x}, \qquad (20)$$

and

$$y_c = v_c + \bar{y}. \qquad (21)$$

From an observed solar image, the two central coordinates and its radius are obtained uniquely.

## Appendix B Correction for Distortion of the Sun's Image

Telescope objective lens with a short focal length as in Kodaikanal's observatory can contribute to a distortion of the projected image. This distortion was corrected by using the following empirical relations:

$$T = \frac{\text{Rad}}{15},$$

$$R_o = 29.5953 \cos\left[\frac{\cos^{-1}(-0.00629T)}{3} + 240\right] \qquad (22)$$

$$\rho' = R_o \times \frac{r}{R}$$

and

$$\rho = \sin^{-1}\left(\frac{\sin(\rho')}{\sin(R_o)}\right) - \rho' \qquad (23)$$

where $T$ is the number of Julian centuries since epoch 1900 January 0.5 and the semidiameter of the Sun, Rad in arcseconds is

$$\text{Rad} = \left(\frac{0.2666}{R'}\right)^0 \times 3600'' \qquad (24)$$

$$R' = 1.00014 - 0.01671\cos(g) - 0.00014\cos(2g)$$

$g$ being the mean anomaly (computed each day).

**ORCID iDs**
K. M. Hiremath  https://orcid.org/0000-0003-0521-9645
J. P. Rozelot  https://orcid.org/0000-0002-5369-1381
A. Kilcik  https://orcid.org/0000-0002-0094-1762
Shashanka R. Gurumath https://orcid.org/0000-0002- 7978-8245